\documentclass[aip,apl,reprint]{revtex4-1}
\usepackage{graphicx}
\usepackage{amsmath}
\begin{document}

\title{Learning Similarity-Reduced Ion Kinetics across Scales and Regimes in Capacitively Coupled Plasmas}

\author{Zeduan Zhang}
\affiliation{School of Physics, Beijing Institute of Technology, Beijing 100081, People's Republic of China}

\author{Jianxiong Yao}
\affiliation{School of Physics, Beijing Institute of Technology, Beijing 100081, People's Republic of China}

\author{Long Chen}
\affiliation{School of Science, Dalian Maritime University, Dalian 116026, People's Republic of China}

\author{Yangyang Fu}
\affiliation{Department of Electrical Engineering, Tsinghua University, Beijing 100084, People's Republic of China}
\affiliation{State Key Laboratory of Power System Operation and Control, Department of Electrical Engineering, Tsinghua University, Beijing 100084, People's Republic of China}
\affiliation{Sichuan Energy Internet Research Institute, Tsinghua University, Chengdu 610213, People's Republic of China}

\author{Feng He}
\affiliation{School of Physics, Beijing Institute of Technology, Beijing 100081, People's Republic of China}

\author{Jiting Ouyang}
\affiliation{School of Physics, Beijing Institute of Technology, Beijing 100081, People's Republic of China}

 \author{Bocong Zheng}
\email{zheng@bit.edu.cn}
\affiliation{School of Physics, Beijing Institute of Technology, Beijing 100081, People's Republic of China}

\date{\today}

\begin{abstract}
Systematic kinetic investigations of discharge-mode and heating-mechanism transitions in low-pressure capacitively coupled plasmas (CCPs) are computationally expensive because particle-in-cell/Monte Carlo collision (PIC/MCC) simulations must resolve the inherently slow ion relaxation.
In this Letter, we propose a similarity-reduced ion-kinetic representation and a machine learning (ML) reconstruction method for weakly ionized Boltzmann-Poisson discharge systems.
Using discharge similarity, the three-dimensional control parameter space, comprising gas pressure $p$, gap distance $d$, and driving frequency $f_\text{rf}$, is reduced into a two-dimensional similarity space $(pd, f_\text{rf}d)$.
A dual-branch ML model predicts the ion density profile at the reference scale and the conditional ion velocity distribution.
Taking a 1D3V argon discharge as the paradigm, we use the reconstructed ion kinetic state to initialize the PIC/MCC simulation, which then relaxes to a self-consistent periodic steady state after only a brief evolution.
Compared to conventional uniform initializations, this approach achieves a maximum computational speedup of 56.8$\times$ and a median speedup of 27.7$\times$.
Furthermore, the physical regimes of complex nonlinear discharge transitions, including the $\alpha$--$\gamma$ mode transition, local-to-nonlocal electron collisional heating mode transition and low-density jumps, are accurately mapped within the similarity space.
Similarity transformations extend these boundaries across scales, opening a new avenue for rapid kinetic characterization and optimization of radio-frequency plasma sources.
\end{abstract}

\maketitle

Low-pressure capacitively coupled plasmas (CCPs) are widely utilized in modern semiconductor manufacturing and surface treatment processes \cite{lieberman2005textbook, chabert2011physics, Chen_2025, Adamovich_2022}.
Their electron heating mechanisms and ion transport characteristics typically require kinetic investigations using particle-in-cell/Monte Carlo collision (PIC/MCC) simulations \cite{Donko_2021, Turner_2013, Yao_2026}.
Within the control parameter space consisting of gas pressure $p$, gap distance $d$, and driving frequency $f_\text{rf}$, different operating conditions can exhibit nonlinear features such as $\alpha$--$\gamma$ mode transitions \cite{Daksha_2017}, local-to-nonlocal electron collisional heating mode transitions \cite{wilczek_2020}, and low-density jumps \cite{Sharma_2016, Vass_2020, Yao_2025}.
The systematic characterization of these features requires covering a wide range of parameters.

However, different parameter combinations do not always represent independent discharge states.
In weakly ionized Boltzmann-Poisson systems, which low-pressure CCPs typically satisfy, the synergistic scaling of different control parameters can correspond to identical reduced states \cite{Fu2023, Fu_2021, zheng2024apl, yao2025similarity}.
Therefore, it is necessary to distinguish between the repetitive descriptions of the same reduced state at different scales and the non-trivial kinetic variations across different discharge regimes.
The redundant information of the former can be analytically eliminated via similarity scaling, whereas the latter still requires solving for the corresponding kinetic states within the reduced parameter space.
Furthermore, reaching a periodic steady state in PIC/MCC simulations is often bottlenecked by inherently slow ion relaxation.
Consequently, we select the kinetic state of the ion as the prediction target to construct a near-steady state ion initial condition.
During the subsequent PIC/MCC evolution, the electron distribution and electric field adjust self-consistently, while residual deviations in the predicted ion distribution relax.

By coupling the similarity reduction of the parameter space with ion kinetic initialization, this Letter proposes a machine learning (ML)-assisted kinetic reconstruction method oriented toward diverse physical scales and discharge regimes.
Taking a 1D3V argon CCP as a paradigm, the three-dimensional (3D) physical control parameter space (hereafter referred to as the parameter space) composed of $p$, $d$, and $f_\text{rf}$ is reversibly decomposed into a two-dimensional (2D) similarity space spanned by the invariants $pd$ and $f_\text{rf}d$, along with an independent scaling factor $\kappa$.
On this basis, the steady-state ion distribution function is factorized into a scale factor, a base-scale density profile, and a conditional ion velocity distribution function (IVDF).
By employing ML to predict these distribution profiles within the dimensionally reduced similarity space, the ion kinetic states at varying scales are reconstructed, providing high-fidelity, near-steady-state initial conditions for simulations.
After a brief period of PIC/MCC computational correction, this initial state rapidly converges to a self-consistent steady state, thereby enabling fast and reliable characterization of kinetic features across the entire parameter space.

Consider the electrostatic Vlasov-Poisson kinetic system governed by
\begin{equation}
\frac{\partial f_s}{\partial t} + \boldsymbol{v} \cdot \nabla_{\boldsymbol{x}} f_s + \frac{q_s}{m_s} \boldsymbol{E} \cdot \nabla_{\boldsymbol{v}} f_s = 0,
\end{equation}
\begin{equation}
    \nabla_{\boldsymbol{x}} \cdot \boldsymbol{E} = \frac{1}{\varepsilon_0}\sum_{s = \text{i,e}}{q_s \int f_s \, \mathrm{d}^3 \boldsymbol{v}},
\end{equation}
where $q_s$ and $m_s$ denote the charge and mass of the species $s$, respectively, and $\varepsilon_0$ is the vacuum permittivity.
To investigate the scaling invariance of this system, we introduce undetermined scaling factors for the spatial, temporal, and velocity coordinates ($\boldsymbol{x}$, $t$, and $\boldsymbol{v}$), the distribution functions $f_s$, and the self-consistent electric field $\boldsymbol{E}$, requiring that any scaled solution identically satisfies the governing equations.
The mathematical invariance of the Vlasov equation necessitates equal scaling factors for its three terms, yielding two independent constraints.
The Poisson equation imposes an additional constraint, leaving the Vlasov-Poisson system with two independent scaling degrees of freedom.
By selecting linear transformations for space and time, $\boldsymbol{x}' = \lambda \boldsymbol{x}$ and $t' = \mu t$, the scaling relations for the remaining variables are strictly determined as $\boldsymbol{v}' = (\lambda / \mu) \boldsymbol{v}$, $\boldsymbol{E}' = (\lambda / \mu^2) \boldsymbol{E}$, and $f'_s(\boldsymbol{x}', \boldsymbol{v}', t') = (\mu / \lambda^3) f_s(\boldsymbol{x}, \boldsymbol{v}, t)$.

To describe the low-pressure CCP discharges considered herein, the Vlasov equation is typically replaced by the Boltzmann equation, featuring a binary collision term $\mathcal{C}_s(f_\text{e}, f_\text{i}, f_\text{n}, \sigma(\varepsilon))$ between charged and neutral particles on the right-hand side.
By assuming negligible neutral gas depletion due to ionization and excitation, the weakly ionized Boltzmann-Poisson system is established.
The inclusion of the collision term introduces a scalable neutral particle distribution function, $f_\text{n}$, along with the constraint that this collision term must scale identically to the other terms.
Furthermore, the relative collision energy scales as $\varepsilon'= (\lambda / \mu)^2 \varepsilon$.
For a given collision reaction, the fixed excitation/ionization thresholds and the inherent energy dependence of the cross sections $\sigma(\varepsilon)$ require the collision energy to remain invariant under scaling, strictly dictating that $\lambda = \mu$. 

Consequently, under this collision model and the aforementioned assumptions, the weakly ionized Boltzmann-Poisson system retains only a single independent scaling degree of freedom, defined as $\kappa = \lambda^{-1} = \mu^{-1}$.
The scaling relations thus become $\boldsymbol{x}' = \boldsymbol{x} / \kappa$, $t' = t / \kappa$, $\boldsymbol{v}' = \boldsymbol{v}$, $\boldsymbol{E}' = \kappa \boldsymbol{E}$, and $f'_s = \kappa^2 f_s$.
Since the binary collision operator is linear with respect to both $f_s$ and $f_\text{n}$ [i.e., $\mathcal{C}_s(\kappa f_\text{e}, \kappa f_\text{i}, f_\text{n}, \sigma(\varepsilon)) = \mathcal{C}_s(f_\text{e}, f_\text{i}, \kappa f_\text{n}, \sigma(\varepsilon)) = \kappa \mathcal{C}_s(f_\text{e}, f_\text{i}, f_\text{n}, \sigma(\varepsilon))$], the scaling constraint on the collision term rigorously yields $f_\text{n}' = \kappa f_\text{n}$.
Provided that the geometry, driving parameters, and boundary conditions satisfy the corresponding transformations, the scaling relations for the macroscopic variables are straightforwardly derived as $d' = d / \kappa$, $f'_\text{rf} = \kappa f_\text{rf}$, $p' = \kappa p$, and $n'_s = \kappa^2 n_s$ \cite{Fu2023, Fu_2021, zheng2024apl}.

\begin{figure}[b]
\includegraphics[width=\columnwidth]{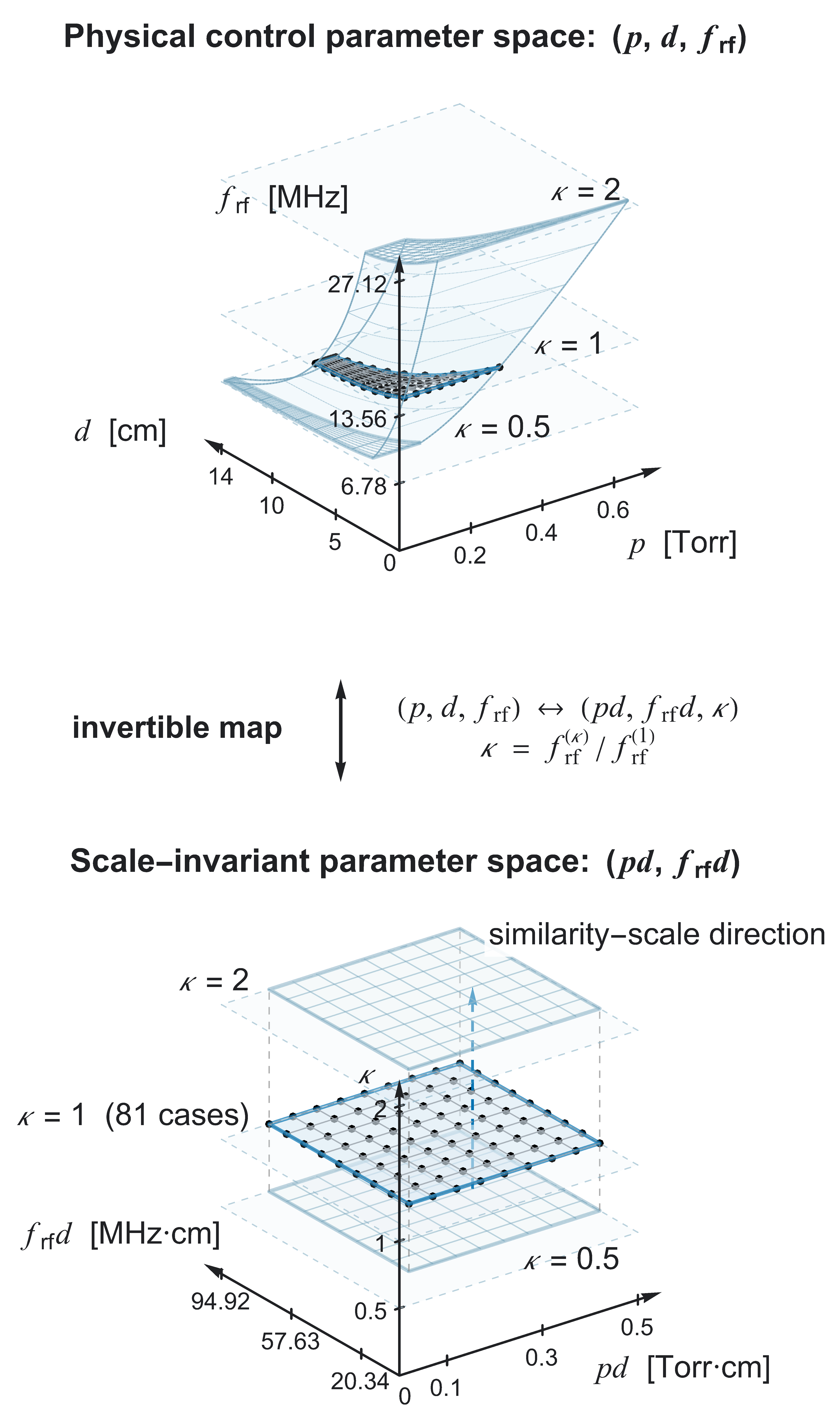}
\caption{\label{fig:map} Mapping from the 3D physical control parameter space $(p, d, f_\text{rf})$ to the 2D scale-invariant parameter space $(pd, f_\text{rf}d)$ and the scaling factor $\kappa$.
The 81 training cases (black dots) are uniformly sampled on the base similarity plane at $\kappa=1$.}
\end{figure}

Following the aforementioned scaling relations, any set of equivalent states within the parameter space can be derived by scaling a single reference state.
By combining the control parameters into the similarity invariants $pd$ and $f_\text{rf} d$, any $(p, d, f_\text{rf})$ configuration sharing identical invariants rigorously corresponds to a similar discharge state. 
Consequently, as illustrated in Fig.~\ref{fig:map}, the 3D parameter space $(p, d, f_\text{rf})$ can be reversibly mapped into a 2D similarity space spanned by $(pd, f_\text{rf} d)$, supplemented by an independent scaling factor $\kappa = f_\text{rf}^{(\kappa)}/f_\text{rf}^{(1)}$ (where $f_\text{rf}^{(1)}$ denotes the base frequency at $\kappa = 1$). 
This mapping cleanly decouples the non-trivial kinetic variations across distinct discharge regimes from the redundant scale dependencies along the $\kappa$-axis.
Employing the scale-invariant phase-space coordinates $\xi \equiv x/d$ and $u \equiv v_x$ (where $v_x$ is the ion velocity along the electrode gap), the steady-state ion distribution function $f_\text{i}^{(\kappa)}(\xi, u)$ is rigorously factorized into the product of three terms:
\begin{equation}
    f_\text{i}^{(\kappa)}(\xi, u) =\kappa^2 n_\text{i}^{(1)}(\xi) \psi_\text{i}(u\vert{}\xi), \ \int\psi_\text{i}(u\vert{}\xi)du = 1
\end{equation}
where the scaling factor $\kappa$ is analytically determined by the driving frequencies of the target and reference conditions.
Conversely, the base-scale ion density profile $n_\text{i}^{(1)}(\xi) = n^{(\kappa)}_\text{i}(\xi) / \kappa^2$ and the conditional ion velocity distribution function (conditional IVDF) $\psi_\text{i}(u\vert{}\xi)$ act as scale-invariant physical quantities along the $\kappa$-axis.
As such, they are learned and predicted by the machine learning model exclusively within the $(pd, f_\text{rf}d)$ similarity space.

\begin{figure}[htbp]
\includegraphics[width=\columnwidth]{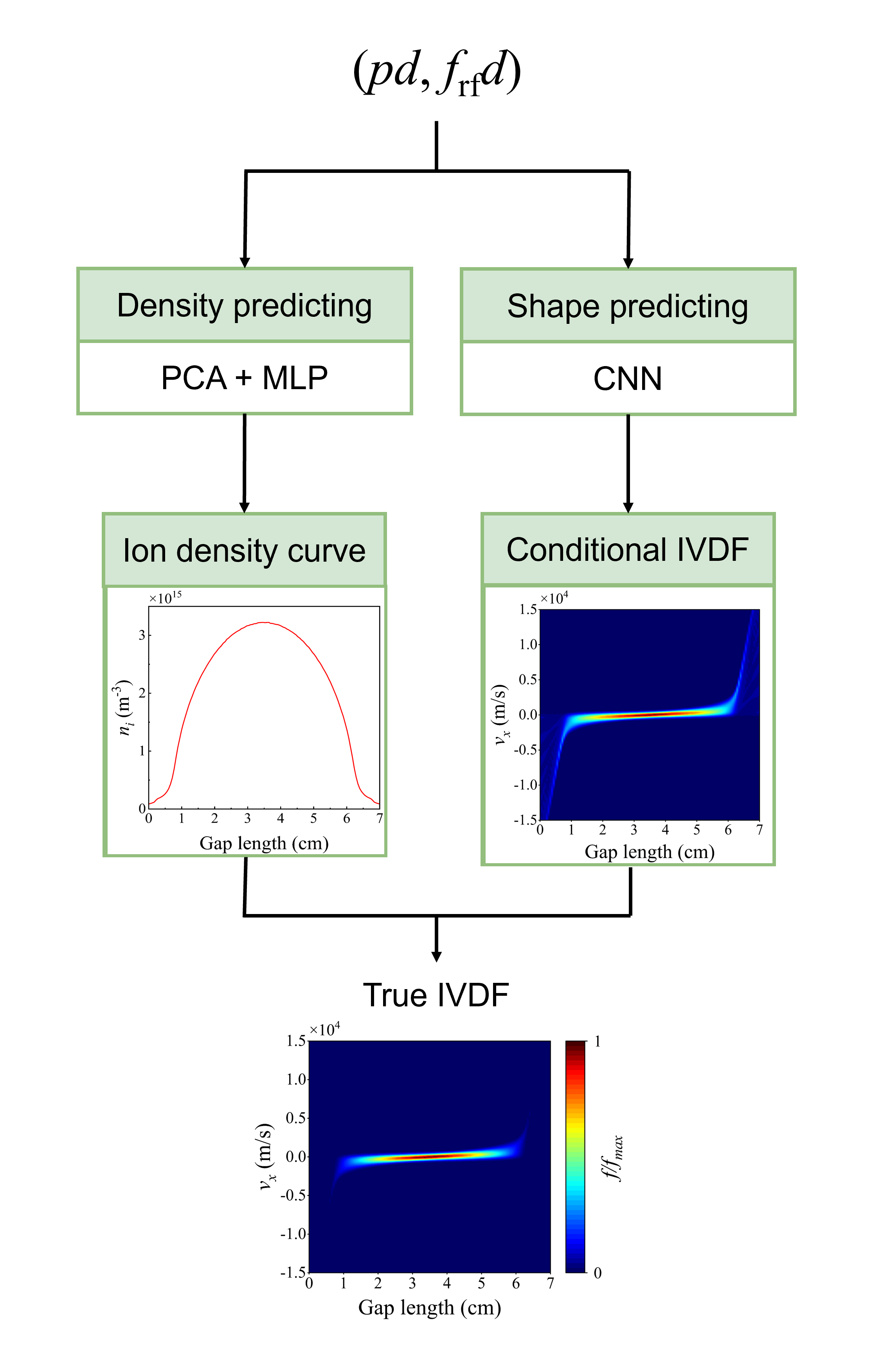}
\caption{\label{fig:workflow} Schematic of the dual-branch machine learning architecture for ion-kinetic state reconstruction.
The scale-invariant parameters $(pd, f_\text{rf}d)$ are fed into parallel networks: a CNN for predicting the conditional IVDF (shape predicting) and a PCA+MLP model for the spatial ion density curve (density predicting).
The outputs from both branches are subsequently recombined to recover the true unnormalized IVDF.}
\end{figure}

\begin{table}[htbp]
\caption{\label{tab:architecture} Architecture and hyperparameters of the dual-branch machine learning model.}
\begin{ruledtabular}
\begin{tabular}{llc}
\textbf{Module} & \textbf{Component / Parameter} & \textbf{Value / Setup} \\
\hline
\multicolumn{3}{c}{\textit{Density Predicting Branch (PCA+MLP)}} \\
\hline
PCA & Principal components & 5 \\ 
MLP & Hidden layers & [32, 64, 32] \\ 
& Activation function & ELU \\ 
Training & Optimizer & Adam\\
& Learning rate & $2 \times 10^{-3}$ \\
& Batch size & 16 \\
& Loss function & L1 Loss\\
\hline
\multicolumn{3}{c}{\textit{Shape Predicting Branch (CNN)}} \\
\hline
Dense & Fully connected layers & [64, 256, 1024, \\
& & 2048, 2048, 1024] \\ 
Conv Blocks & ConvTranspose2D & (16, $4\times4$), (8, $4\times4$), \\ 
& (filters, kernel) & (4, $4\times4$), (1, $4\times4$) \\ 
Pooling & Down/Up-sampling & Stride-2 upsampling \\ 
& strategy & (No pooling) \\ 
& Activation function & ReLU (hidden), \\
& & LeakyReLU (output) \\ 
Training & Optimizer & Adam \\ 
& Learning rate & $1 \times 10^{-3}$ \\ 
& Batch size & 32 \\ 
& Loss function & MSE Loss\\ 
\end{tabular}
\end{ruledtabular}
\end{table}

To accurately capture these two physical quantities with distinct distribution characteristics, a dual-branch machine learning model is designed for joint learning and reconstruction, as illustrated in Fig.~\ref{fig:workflow} (detailed hyperparameters are provided in Table~\ref{tab:architecture}).
In the density-predicting branch, since the steady-state $n_\text{i}^{(1)}(\xi)$ is a smooth one-dimensional macroscopic spatial profile devoid of sharp local variations, we employ principal component analysis (PCA) coupled with a multilayer perceptron (MLP) for dimensionality reduction and feature prediction.
The steady-state spatial density distribution is ultimately recovered via an inverse PCA transform \cite{Powis_2026}. 
Conversely, in the shape-predicting branch, $\psi_\text{i}(u\vert{}\xi)$ encapsulates intricate 2D phase-space local kinetic features.
Therefore, a convolutional neural network (CNN) is employed to predict its distribution feature maps \cite{Powis_2026, Faraji_2025, Bonzanini_2023}, followed by deconvolutional operations to reconstruct the localized probability distributions at specific spatial positions. 
To construct the training dataset, 81 independent operating conditions are uniformly sampled within the base similarity plane (Fig.~\ref{fig:map}, $\kappa=1$).
Across all simulations, the discharge voltage is fixed at 200 V, and the secondary electron emission coefficient is set to 0.2.
Upon training completion, the predictive outputs from both branches are recombined with the analytical scale factor $\kappa^2$, enabling the rapid reconstruction of the complete, dimensioned true ion velocity distribution function (true IVDF). 
The reconstructed kinetic state is subsequently injected as the initial condition for PIC/MCC simulations, rapidly eliminating residual errors and achieving a self-consistent physical steady state.
All particle simulation data utilized in this Letter are generated by our in-house 1D3V electrostatic PIC/MCC code, ASTRA \cite{Zheng_2019, zheng2024apl, yao2025similarity, Yao_2025, Yao_2026}.

\begin{figure}[b]
\includegraphics[width=\columnwidth]{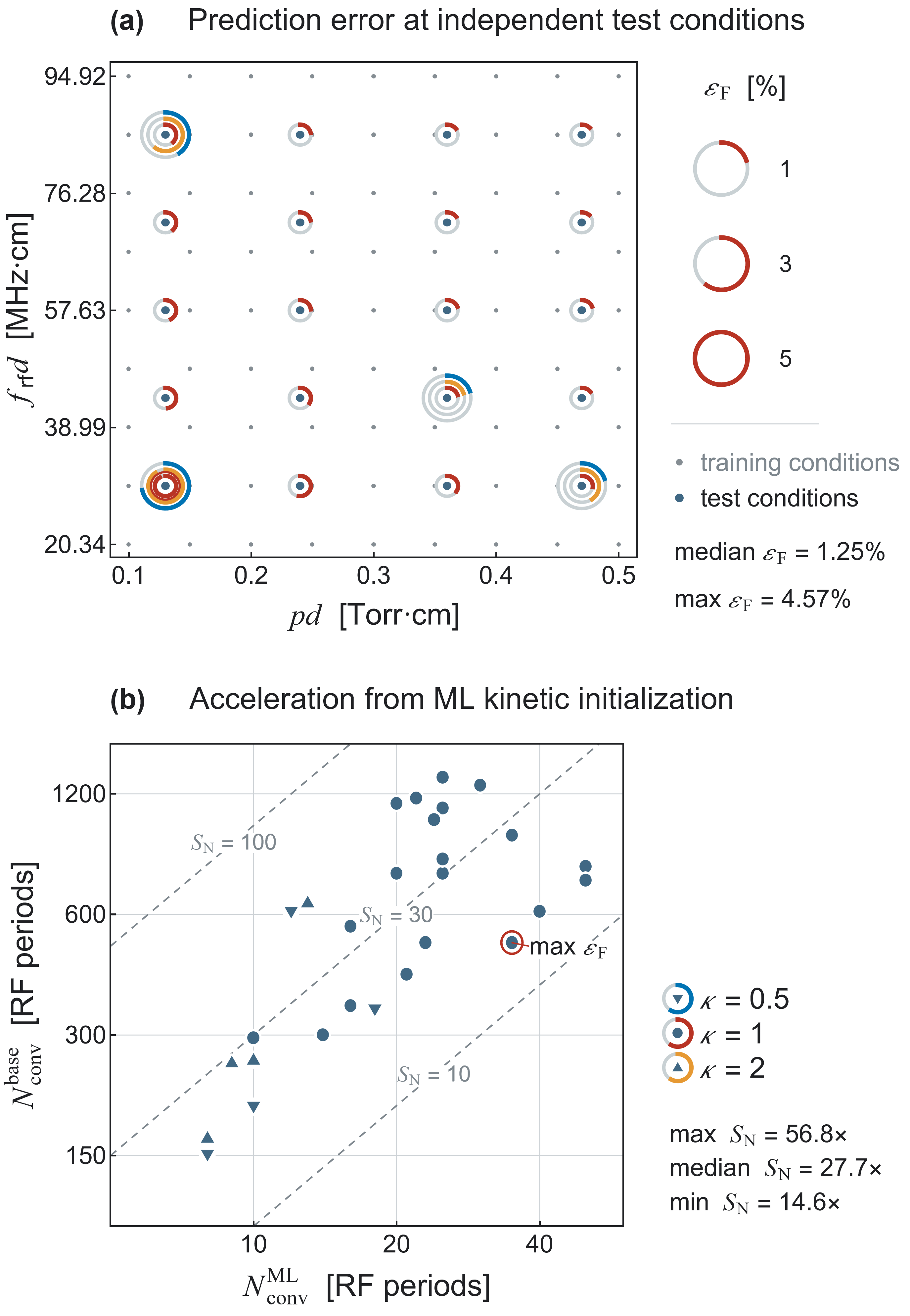}
\caption{\label{fig:result} Performance evaluation of the ML-based kinetic reconstruction.
(a) Prediction errors $\epsilon_f$ of the reconstructed IVDFs for independent test conditions in the similarity space.
The colored arc fractions denote the error magnitude.
(b) Computational acceleration achieved by the ML kinetic initialization.
The scatter plot compares the required RF periods to reach steady state using the traditional homogeneous ($N_{\text{conv}}^{\text{base}}$) and ML-based ($N_{\text{conv}}^{\text{ML}}$) approaches.
The scaling factors are represented by concentric ring colors in panel (a) and marker shapes in panel (b): $\kappa=0.5$ (blue, down-triangles), $\kappa=1$ (red, circles), and $\kappa=2$ (orange, up-triangles).
Dashed lines indicate contours of constant speedup $S_N$.}
\end{figure}

To validate the aforementioned similarity-reduced kinetic representation and the generalization capability of the dual-branch ML model, its predictive accuracy and computational speedup are systematically evaluated on an independent test set, as depicted in Fig.~\ref{fig:result}. 
Figure~\ref{fig:result}(a) illustrates the model's prediction error, $\epsilon_f$, for the IVDF across unseen operating conditions within the similarity space.
Notably, although the model is exclusively trained on the base similarity plane ($\kappa=1$), the test dataset encompasses varied scaling factors ($\kappa=0.5, 1, \text{and } 2$) to rigorously verify its extrapolation capability across macroscopic physical scales. 
The evaluation demonstrates high fidelity across all independent conditions, yielding a median prediction error of merely $1.25\%$.
Even under the most extreme boundary conditions, the maximum error remains strictly bounded at $4.57\%$. 
This directly confirms that the proposed similarity-reduction strategy successfully captures the intrinsic kinetic invariants governing the inherently slow ion relaxation processes.

Building upon this highly efficient and high-fidelity kinetic reconstruction, Fig.~\ref{fig:result}(b) further illustrates the immense acceleration potential of the proposed method in practical PIC/MCC simulations. 
Traditional PIC/MCC simulations, initialized with uniform density profiles, typically require thousands of radio-frequency (rf) cycles ($N_{\text{conv}}^{\text{base}}$) to resolve the inherently slow ion relaxation processes before establishing a self-consistent periodic steady state. 
In contrast, by injecting the ML-reconstructed ion kinetic state as the simulation initial condition, the number of evolution cycles required to reach the identical convergence criterion \cite{Powis_2026} ($N_{\text{conv}}^{\text{ML}}$) is drastically reduced. 
The evaluation indicates that this approach achieves a remarkable computational speedup ($S_N$) across the entire parameter space, yielding a median of $27.7\times$ and a maximum of up to $56.8\times$.
By directly bypassing the most time-consuming ion relaxation stage, this ML-driven kinetic initialization strategy, coupled with the similarity-based dimensionality reduction, renders systematic kinetic parameter sweeps across high-dimensional spaces—previously considered computationally prohibitive—entirely feasible.

\begin{figure}
\includegraphics[width=\columnwidth]{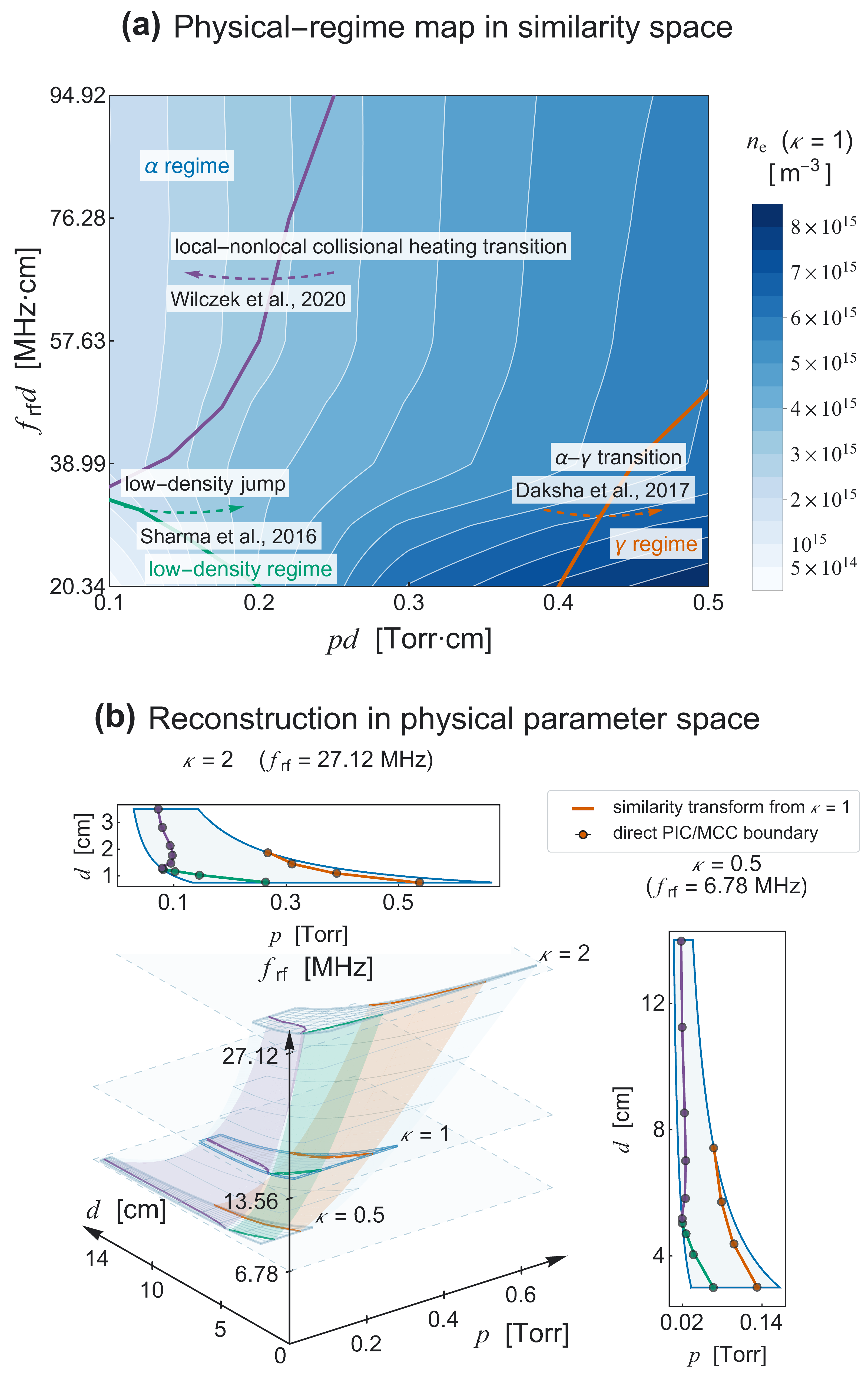}
\caption{\label{fig:regime} Mapping of physical regimes and their cross-scale reconstruction.
(a) Physical-regime map in the scale-invariant $(pd, f_\text{rf}d)$ space, distinguishing various nonlinear discharge transitions (e.g., $\alpha$--$\gamma$ transition, local-nonlocal collisional heating transition, low-density jump).
The background contours represent the electron density at $\kappa=1$.
(b) Reconstruction of the topological boundaries back into the 3D physical $(p, d, f_\text{rf})$ space.
The similarity predictions scaled from the $\kappa=1$ base plane (solid lines) are perfectly verified by direct PIC/MCC calculation boundaries (symbols) at $\kappa=0.5$ ($f_\text{rf}=6.78\text{ MHz}$) and $\kappa=2$ ($f_\text{rf}=27.12\text{ MHz}$).}
\end{figure}

Having verified the reconstruction accuracy and computational speedup of the model, we further investigate the distribution of nonlinear discharge mode transitions within the similarity space, as depicted in Fig.~\ref{fig:regime}(a). 
Empowered by the highly efficient ML-driven kinetic predictions, we can systematically map the contours of the time-averaged electron density across the continuous $(pd, f_\text{rf}d)$ similarity space, precisely delineating the boundaries of various complex discharge transitions inherent to low-pressure CCPs. 
Alongside the conventional $\alpha$ and $\gamma$ discharge modes and their nonlinear boundary ($\alpha$--$\gamma$ transition) \cite{Daksha_2017}, we clearly capture the low-density jumps in the low-pressure regime \cite{Sharma_2016, Vass_2020}, as well as the demarcation for the local-to-nonlocal electron collisional heating transition \cite{wilczek_2020}. 
Although these highly nonlinear macroscopic discharge transitions typically violate traditional simple linear scaling laws, they exhibit unified and distinct demarcations within the 2D similarity space across varying physical scales.

This similarity-reduced representation exhibits rigorous reversibility and cross-scale reconstruction capabilities, as illustrated in Fig.~\ref{fig:regime}(b).
Utilizing the independent scaling factor $\kappa$, the complex discharge transition boundaries resolved within the similarity space are successfully mapped back into the real 3D physical parameter space $(p, d, f_\text{rf})$. 
To verify the reliability of this cross-scale reconstruction, independent PIC/MCC simulations are executed on the parameter cross-sections at $\kappa=0.5$ (corresponding to $f_\text{rf}=6.78\text{ MHz}$) and $\kappa=2$ (corresponding to $f_\text{rf}=27.12\text{ MHz}$).
The comparative results demonstrate an excellent agreement between the discharge boundaries reconstructed purely via system similarity and ML scaling (solid lines) and those obtained directly from simulations (scatter symbols). 
This not only confirms that the steady-state IVDF in CCPs serves as a kinetic representation strictly satisfying similarity transformations, but also proves that the framework integrating discharge similarity theory with data-driven methods possesses a robust cross-scale extrapolation generalization capability when handling complex nonlinear plasma evolutions.

In summary, this Letter develops and evaluates an ion kinetic reconstruction method coupling parameter similarity reduction with machine learning for low-pressure capacitively coupled plasmas.
By isolating the redundant scale dependencies across diverse operating conditions via discharge similarity, the three-dimensional parameter space is dimensionally reduced into a two-dimensional similarity space.
A dual-branch machine learning model is then employed to learn and predict this kinetic representation, enabling the complete self-consistent kinetic state to rapidly converge through a brief subsequent PIC/MCC evolution. 
The findings verify that this similarity-reduced kinetic representation reproduces the complex nonlinear discharge evolution regimes and macroscopic physical scalings with high fidelity.
Across all tested operating conditions and scales, this initialization strategy preserves the accuracy of the ion kinetic state reconstruction while significantly accelerating the PIC/MCC computations to meet the steady-state convergence criteria. 
Ultimately, this methodology renders systematic kinetic parameter sweeps across vast parameter spaces—previously considered computationally prohibitive—entirely feasible.
It opens a highly efficient new avenue for the rapid and reliable characterization of kinetic properties in various plasma systems, including radio-frequency sources, and holds profound promise for providing high-fidelity data backbones for the construction of plasma equipment digital twins.

\begin{acknowledgments}
This work was partly supported by the National Natural Science Foundation of China (No. 12675254 and 11975047).
\end{acknowledgments}

\section*{Data Availability Statement}
The data that support the findings of this study are available from the corresponding author upon reasonable request.
 
\bibliography{Reference}

\end{document}